\documentclass[conference]{IEEEtran}

\usepackage{graphicx}
\usepackage{fancyhdr}
\usepackage{color}
\usepackage{xspace}
\usepackage{amsmath}
\usepackage{url}
\usepackage{latexsym}
\usepackage{amssymb}
\usepackage{subfigure}
\usepackage{flushend}
\usepackage{tabularx}
\usepackage{mathtools}
\usepackage{caption}
\usepackage{varwidth}
\usepackage{comment}
\usepackage{fixltx2e}
\usepackage[usenames,dvipsnames]{xcolor}

\newif\ifFullVersion
\FullVersiontrue

\newif\ifShortVersion
\ifFullVersion
  \ShortVersionfalse
\else
  \ShortVersiontrue
\fi

\ifShortVersion
  \excludecomment{fullonly}
\fi

\ifFullVersion
  \excludecomment{shortonly}
\fi

\usepackage[noadjust]{cite}

\renewcommand{\theenumi}{\roman{enumi}.}

\makeatletter
\renewcommand*{\thetable}{\arabic{table}}
\renewcommand*{\thefigure}{\arabic{figure}}
\let\c@table\c@figure
\makeatother

\renewcommand\tablename{Fig.}
\usepackage{algorithm}
\usepackage[noend]{algpseudocode}
\DeclareCaptionFormat{myformat}{#3}
\algnewcommand{\LineComment}[1]{\State \(\triangleright\) #1}
\usepackage[normalem]{ulem}

\definecolor{lightblue}{rgb}{0.73,0.73,0.73} % color values Red, Green, Blue

\usepackage[breaklinks]{hyperref}       %<-- these fix url line breaks
\usepackage{breakurl}   %<-- you need to compile twice the 1st time

\expandafter\def\expandafter\UrlBreaks\expandafter{\UrlBreaks%  save the current one
  \do\a\do\b\do\c\do\d\do\e\do\f\do\g\do\h\do\i\do\j%
  \do\k\do\l\do\m\do\n\do\o\do\p\do\q\do\r\do\s\do\t%
  \do\u\do\v\do\w\do\x\do\y\do\z\do\A\do\B\do\C\do\D%
  \do\E\do\F\do\G\do\H\do\I\do\J\do\K\do\L\do\M\do\N%
  \do\O\do\P\do\Q\do\R\do\S\do\T\do\U\do\V\do\W\do\X%
  \do\Y\do\Z}

\newcommand{\tinyskip}{\vspace{0.5pt}}
\newcommand{\sskip}{\vspace{1pt}}
\newcommand{\name}{FAIR\xspace}
\newcommand{\nam}{fair\xspace}
 
\newcommand{\block}[1]{}                % Comment out large swaths of text

\def \R { \overset{R}{\leftarrow} }   % Choose at random

\newcommand{\hash}[1]{
	\text{H(}#1\text{)}
}
\newcommand{\sign}[2]{
	\text{Sign}_{#1}\text{(}#2\text{)}
}

\renewcommand\paragraph[1]{\smallskip\noindent\textbf{#1.}}

\begin{document}

\title{\huge\name: Forwarding Accountability for Internet Reputability}
\author{
{\rm Christos Pappas}\\
{\small ETH Z\"urich}\\
pappasch@inf.ethz.ch
\and
{\rm Raphael M. Reischuk}\\
{\small ETH Z\"urich}\\
reischuk@inf.ethz.ch
\and
{\rm Adrian Perrig}\\
{\small ETH Z\"urich}\\
adrian.perrig@inf.ethz.ch
}
\maketitle

\begin{abstract}
This paper presents \name, a forwarding accountability
mechanism that incentivizes ISPs to apply stricter security policies to their
customers. The Autonomous System (AS) of the receiver specifies a
traffic profile that the sender AS must adhere to. Transit ASes on the path
mark packets. In case of traffic profile violations, the marked packets are
used as a proof of misbehavior.

\name introduces low bandwidth overhead and requires no per-packet
and no per-flow state for forwarding.
We describe integration with IP and demonstrate a
software switch running on commodity hardware that can
switch packets at a line rate of 120~Gbps, and can forward
140M minimum-sized packets per second, limited by the
hardware I/O subsystem.
%
%Furthermore, \name is compatible with IP networks.

Moreover, this paper proposes a ``suspicious bit'' for
packet headers --- an application that builds on top of
\name's proofs of misbehavior and flags packets to warn
other entities in the network. 

\end{abstract}

\section{Introduction}
\label{sec:intro}
The frequency and intensity of attacks rooted in misconfigured or 
vulnerable Internet services has increased in the last months: in February
2014, attackers abused misconfigured time synchronization
servers~\cite{ntp_attack} to attack Cloudflare with a peak of
400~Gbps~\cite{cloudflare}.
For 2014, Akamai reports a 90\% increase in total DDoS attacks and a 52\%
increase in average peak bandwidth compared to the previous
year~\cite{akamai_q4}. Moreover, man-on-the-side script injection
attacks~\cite{baidu} and vulnerable web services have been used as general-purpose 
attack vectors~\cite{itsoknoprob,drupal}.

These events are explained by the following observations.  First, the lack of
accountability in today's Internet facilitates attacks with spoofed addresses,
allowing attackers to evade blocking mechanisms.  Second, the architectural
limitations of today's Internet lead to insufficiently effective DDoS defense
mechanisms.  Third, ISPs have no incentive to punish their misbehaving
customers, nor to monitor them. Typically, monitoring comes with high storage
and computational requirements that yield additional costs for network
operators.

In order to address these problems, the security community has considered
several solutions, which come with certain shortcomings: \textit{source
accountability schemes}~\cite{aip,passport} encounter routing scalability
problems and introduce prohibitive bandwidth overhead; \textit{cloud-based
retroactive DDoS defense services} introduce latency and are insufficiently
effective, yet prices can exceed several thousand dollars per
month~\cite{cloudflare_prod}; \textit{capability
schemes}~\cite{siff,tva,portcullis,capabilities} introduce complexity and
require per-flow operations; \textit{extensive
filtering}~\cite{stopit,aitf,pushback} requires operator vigilance and
out-of-band coordination among ISPs.

Although we stand in solidarity with these proposals, this paper takes a
different approach and proposes a lightweight scheme that incentivizes ASes to solve their security problems. 
To this end, we leverage \textit{forwarding accountability}.
In a nutshell, the key idea behind forwarding accountability is to hold ASes
accountable for the traffic they forward; transit ASes embed proofs in the
packets such that, in case of malicious traffic, a destination AS can later use
these proofs to show to the transit ASes that they have indeed forwarded the
malicious traffic. We stress that transit ASes do not store any information, but given proofs
of misbehavior they can deprioritize traffic from provably malicious ASes.
This protects the victim and increases capacity for benign traffic.

We take volumetric DDoS attacks as one possible use case and demonstrate the
virtues of forwarding accountability.  Consider the topology depicted in
\autoref{fig:networkmodel} and assume web servers, or even servers of critical
infrastructures, are located inside AS\textsubscript{n}. We assume, exactly as
happened in 2014~\cite{cloudflare}, that an attacker launches a reflection
attack against the victims by exploiting the NTP protocol running on
misconfigured servers. More precisely, the attacker fakes the victim's source
IP address and sends NTP commands to the misconfigured NTP servers within
AS\textsubscript{0}.  The NTP servers reply to the victim with responses that
are up to 200 times larger than the initial rogue requests, overpowering the
victim's resources. With forwarding accountability in place, the transit ASes
embed proofs in the packets that will remind them later that they forwarded the
traffic. When the victim reports the attack to transit ASes
(AS\textsubscript{1} and AS\textsubscript{2}) by providing the proof, the
transit ASes acknowledge that they indeed forwarded the malicious traffic.  It
then becomes evident that AS\textsubscript{0} sourced the malicious traffic,
namely from the misconfigured NTP servers. AS\textsubscript{1} can then drop
(or at least deprioritize) AS\textsubscript{0}'s traffic and thus protect not
only AS\textsubscript{n} and its servers, but also all networks between
AS\textsubscript{0} and AS\textsubscript{n}. This approach provides benefits
also in sparse deployment, where only one transit AS accepts proofs of
misbehavior and takes action. Hence, adoption does not require coordination
among ISPs.

A cost-effective incremental deployment path is critical to the success of any
practical security scheme. ISPs' willingness to adopt security mechanisms is
motivated by their reputation and the competitive market environment~\cite{rmanifesto}, but
constrained by the additional expenses and the lack of economic
incentives~\cite{idsurvey_ccr}. In addition, recent Internet regulations~\cite{open_iorder}
intend to actively involve ISPs in stopping the dissemination of malicious traffic,
thus making security mechanisms a necessity in the near future.
Despite regulatory pressure for adoption of security mechanisms, 
efficiency and incremental deployment remain important properties
that drive adoption.

\sskip
\paragraph{Contributions}
This paper proposes an architectural mechanism, \name, to achieve Forwarding
Accountability for Internet Reputability. The key concept is that transit
ASes embed short cryptographic markings in the packets that will later prove
to the ASes that they forwarded these packets. In case of malicious traffic,
destination ASes can use these proofs to show to transit ASes that they
have indeed forwarded the malicious traffic. After acknowledging the proof
of misbehavior, the transit ASes can deprioritize traffic from malicious
ASes, increasing network capacity for benign sources.

\name is founded on a strong threat model where source,
destination, and transit ASes can be compromised or malicious.
Moreover, \name has the following properties:
\begin{itemize} 
  \item low overhead for processing and bandwidth.
  \item no per-packet and no per-flow state for forwarding.
  \ifFullVersion
    \item simple key management with one shared key between source and
    destination ASes.
  \fi
  \item deployment compatibility with IP networks.
  \ifFullVersion
    \item complementary applicability to DDoS defense schemes.
  \fi
\end{itemize}

We have designed and implemented a software switch performing \name
packet marking that operates at line rate
of up to 120~Gbps; it forwards 140M minimum-sized packets per second on a
commodity machine, which is currently limited by the hardware I/O subsystem.

With \name in place, we reconsider Bellovin's April Fool proposal of the
``evil bit''~\cite{rfc3514} and propose an extension to our proposal, the
``suspicious bit'': ASes that forward traffic from misbehaving customers mark
this traffic as suspicious, informing other entities in the network. The
suspicious bit provides a strong incentive for an AS to watch its traffic and
mark malicious traffic itself with the suspicious bit, otherwise its upstream
ISP may mark all of the AS's traffic as suspicious, thus,
causing collateral damage to benign senders.

%\section{Problem Setup}
%\label{sec:problem}
%\input{sec/problem}

\section{Overview of \name}
\label{sec:overview}
\begin{figure}[!t]
  \centering
  \includegraphics[width=.90\columnwidth]{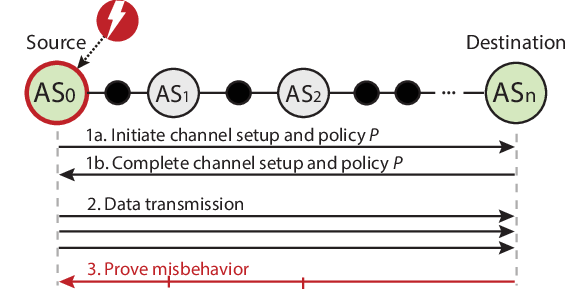}
  \caption{Communication under \name.}
  \label{fig:networkmodel}
\end{figure}

Before describing our assumptions and protocol details, we first present a
high-level overview of \name.  Our proposal combines ideas from capability
systems~\cite{siff, tva, portcullis} and traceback mechanisms~\cite{traceback},
yet its approach is fundamentally different: instead of carrying capabilities,
packets collect proofs that will remind transit ASes of having forwarded these
packets. In case of malicious traffic, the destination AS sends the proofs
\textit{back} to the transit ASes.  Communication under \name proceeds in three
phases. These are depicted in \autoref{fig:networkmodel} using a line-network
topology with cooperating ASes (gray circles) and non-cooperating ASes (black
circles)\footnote{This is a simplified communication model, which assumes that
all flows from the source to the destination AS follow the same AS path. We
will relax this assumption later.}. \textit{Cooperating ASes} are ASes that
support \name, which, however, does not imply benign behavior.

\begin{itemize}
	\item \textbf{Phase 1 (Setup):} Source and destination ASes set up a communication channel
and agree on a sending policy that governs the aggregate traffic from the source AS to the destination
AS over a specific AS path. Such a policy can specify the average sending rate, 
the maximum burst size, or even forbid abnormal packet headers that are used for OS fingerprinting and 
flooding attacks (e.g., Christmas tree packets~\cite{xmas}).

	\item \textbf{Phase 2 (Transmission):} The source sends data
packets to the destination over the communication channel. Each cooperating transit
AS inscribes minimal information in the packet headers, which serves
as a proof to itself that it has forwarded the packets.

	\item \textbf{Phase 3 (Protest):} If the destination AS detects
	a policy violation, it proceeds to the protest phase and provides
	the sending policy together with the data packet headers to the transit ASes, as a proof
	of misbehavior. This proof of misbehavior identifies the adversary.
\end{itemize}

Setting up a sending policy specifies the sending properties of the aggregate
traffic from the source AS to the destination AS.  A violation implies that the
source AS is compromised, malicious, or has poor security practices. A
destination AS, depending on its security policies, can drop traffic from
source ASes that do not set up a sending policy. Transit ASes receive the proof
of misbehavior and can deprioritize inappropriate traffic, depending on their
policies.  In the DDoS use case, a destination AS establishes a traffic profile
with its source ASes and specifies the receiving rates according to its
resources. Hence, in case of an attack, the destination AS can prove to transit
ASes the sending rate violations.

\subsection{Setup (Phase 1)}
\label{sec:phase1}
Source and destination ASes set up a channel with a sending policy $P$ 
for traffic from the source AS to the destination AS.
The sending policy is formally expressed by the Token Bucket (TB)
parameters~\cite{cisco_shape} that the source AS should use for traffic shaping towards the
destination AS. In the TB algorithm, a fixed-sized bucket is filled with tokens
at a certain rate.  A token represents a permission to send a specific number
of bits. For a packet transmission, a number of tokens equal to the packet size
is removed from the bucket. If there are not enough tokens, the packet either
waits for more tokens (shaper) or is discarded (policer).  The TB is the formal
description of the properties of a transmission. It allows burstiness, but
bounds it, as the maximum burst size is proportional to the bucket size. 

More specifically, the destination AS specifies two parameters: the Committed Information Rate
($\mathit{CIR}$), i.e., the average amount of data sent per time
unit; and the Committed Burst Size ($\mathit{CBS}$), i.e., the
maximum amount of data that can be sent (for a given time interval).
The time interval ($\mathit{T_c}$) is determined through
the relation $\mathit{CIR} = \mathit{CBS}/{T_c}$. Using these
values, the sending policy is then established as follows:

\begin{enumerate}
	\item The source AS constructs a sending policy packet and sends it to
the destination AS.
	\item Each cooperating transit AS indicates its presence on
the path. It does not interfere with the sending policy details, nor does it
keep per-policy state.
	\item The destination AS completes the sending policy by filling in the
$\mathit{CIR}$ and $\mathit{CBS}$ values and returns the information
to the source AS.
\end{enumerate}

This is merely an example of a policy construction to demonstrate
the necessary information to prove misbehavior in the data plane, which is our focus.
For example, to handle temporary increased traffic volumes (e.g., during popular
sport events) the source AS can renegotiate the policy's properties and request
more bandwidth.

The setup phase can also be substituted by other future Internet proposals.
For example, Route~Bazaar~\cite{routebazaar} uses publicly verifiable
multilateral contracts among ASes; SCION~\cite{scion_arxiv,scion} provides
explicit path-validation information for AS paths.

\subsection{Data Transmission (Phase 2)}
We describe the data-plane operations performed by source ASes,
cooperating ASes, and destination ASes. These operations are
applied to each data packet.

\sskip
\paragraph{Source AS}
The source sends data packets over
the known path. Border routers of the source AS enforce the sending
policy by applying the parameters to the Token Bucket. Moreover, they
embed additional information in the packet, including a sequence
number and a sending time. This information is used at a later stage
to construct a proof of a violation.

\sskip
\paragraph{Transit ASes}
Each egress border router of a cooperating transit AS performs the
following operations upon packet reception:

\begin{enumerate}

\item The border router verifies that the source's
timestamp in the packet is recent and does not deviate from the
local time beyond a threshold, otherwise the
border router drops the packet.

\item The border router marks the packet, indicating that it has
``seen'' the packet.  The marking is cryptographically protected with a
message authentication code. Since each marking is used to remind
only the corresponding AS that inscribed it, it is computed with a
secret key that is only known to the AS. This marking is used in the
third phase to remind the corresponding AS that it indeed forwarded
the packet.
\end{enumerate}

\sskip
\paragraph{Destination AS}
The destination AS monitors the communication channel and performs traffic
policing to detect sending policy violations.  It stores only packet headers as they
contain the markings for the proof of misbehavior, which enables the corresponding
transit ASes to acknowledge that they indeed forwarded the packets. If a violation
is detected, the destination uses the received packets and proves misbehavior
to the transit ASes.

\subsection{Proving Misbehavior (Phase 3)}
\label{subsec:phase3}
The goal of Phase 3 is to enable destination ASes to provably protest to
other ASes.  Taking action against misbehavior is a decision that a
destination AS makes according to its interests and policies.
Complaints for malicious behavior is an offline procedure between the
destination and the transit ASes.  The procedure occurs in two
rounds. 

First, the destination provides the sending policy and the data packet
headers to all cooperating ASes on the path. The sending policy contains the
transmission properties ($\mathit{CIR}$ and $\mathit{CBS}$) for the
communication channel. The data packet headers contain information
for the actual transmission properties. The ASes examine the evidence
and acknowledge or reject the complaint. An approved complaint means
that the AS acknowledges that it forwarded inappropriate traffic
compared to the sending policy specification. This, however, does not mean
that the source AS is malicious. For example, if a transit AS injects packets,
the source is not responsible for the violation. The destination AS
collects approved and rejected complaints from the transit ASes.

In the second round, the destination AS sends all the collected
information back to the ASes. Based on this information, the ASes on
the path conclude whether the source AS is compromised or whether there
were attempts to falsely blame an innocent source. In
\autoref{sec:security_anal} we explain situations with malicious
transit ASes.

\section{The \name Protocol}
\label{sec:protocol}
We make certain design choices that construct a lightweight
accountability mechanism: 1)~proofs of misbehavior are
carried in data packets, allowing stateless forwarding for
transit ASes; 2)~probabilistic detection of misbehavior
introduces minimal overhead per-packet (a few bytes), keeping
bandwidth overhead low; 3)~all data-plane cryptography is symmetric,
degrading forwarding performance marginally.

\subsection{Assumptions}
\begin{itemize}
	\item \textit{The source knows the AS-level path to the destination
and also knows which ASes on the path deploy the mechanism}. BGP
update messages contain the AS-level path in the AS-path
attribute~\cite{rfc4271} and cooperating ASes can advertise their
support for \name in their BGP announcements as a transitive
attribute.

	\item \textit{Participating parties can obtain and authenticate the public keys of
all cooperating ASes.} We leverage RPKI~\cite{rpki}, a PKI framework that enables
entities to authenticate resource certificates (issued by Regional
Internet Registries) that bind Autonomous System Numbers (ASNs) to
the corresponding public keys, given the correct RPKI public root
key.

	\item \textit{Source and destination ASes perform traffic shaping and
policing based on the Token Bucket algorithm}. For example, Cisco's
shaping mechanisms (Generic Traffic Shaping, Class-Based Shaping, Distributed
Traffic Shaping) and policing mechanisms (Committed Access Rate, Traffic
Policing) are based on the Token Bucket~\cite{cisco_shape}.
\end{itemize}

Furthermore, we assume that the cryptographic mechanisms are secure, i.e.,
cryptographic hash functions cannot be inverted, signatures cannot be forged,
and encryptions cannot be broken.

\subsection{Parameters}
\paragraph{Cryptographic Operations} 
Source and destination ASes establish a secret key ($K_{\mathit{SD}}$) between
them and cache the key to avoid redundant computations. To establish
the key, they can obtain the public keys from the RPKI and use a
non-interactive Diffie-Hellman key exchange~\cite{skip,rfc5246}.
Furthermore, each transit AS\textsubscript{i} uses two local secret
keys that can be changed independently from the other ASes: one
long-term key for control-plane operations ($\hat{K_i}$) and one key
for data-plane operations ($K_i$). These local secret keys are
independent of the communication channels that traverse the AS.
Furthermore, transit ASes keep the previous keys for at least $T_m=12$
hours to be able to verify proof that refers further to the past.

\tinyskip
\paragraph{Protest Time Margin ($T_m$)} 
The destination can protest right after a violation is detected or defer the
process to a later point in time. However, we set a time margin after which
transit ASes are not obliged to examine proofs of violations, to avoid
situations where complaints refer to violations too far in the past. The value
for this parameter is agreed upon and universally known to the cooperating
ASes. There is no strict requirement for choosing this value; we use
$T_m=12$~hours so that ASes have a loose time window to prove misbehavior.

\begin{figure}[t]
	\small
	\centering
	\includegraphics[width=.9\columnwidth]{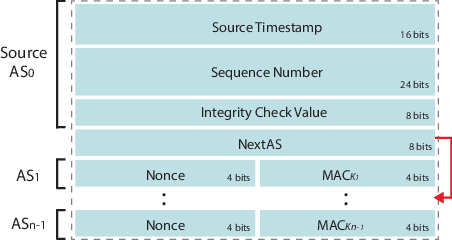}
	\caption{\name Packet Header.}
	\label{fig:acdcheader}
\end{figure}

\tinyskip
\paragraph{Clock Deviation} 
We assume loose clock synchronization between ASes and a reference clock can be set up
with an error less than 0.5 seconds; GPS can provide sub-microsecond 
precision~\cite{gps}. Furthermore, we assume that the end-to-end
packet latency (propagation, transmission, queuing, and processing
delay) does not exceed one second. 

Reported timestamps in packets are at the granularity of seconds, hence
packets with timestamps that differ more than three seconds from the
local time at each router are dropped. This check ensures that the
timestamps in the packets are fresh and can be used in the protest
phase. The three-second margin ensures that packets will not get
dropped due to boundary effects when the end-to-end latency and the
clock difference add up (one second maximum clock difference between 
ASes, one second maximum end-to-end latency, and one second due to possible
boundary effects during clock transitions).

\subsection{Protocol Operations}
We describe the required operations starting with the data plane, which
realizes our notion of forwarding accountability. Then, in the control plane,
we present a low-latency channel setup and the corresponding sending policy
construction. In the end, we show how the sending policy and the data packets
are used to prove misbehavior. \autoref{table:notation} summarizes the notation we use
throughout the paper.

\sskip
\paragraph{1) Data Plane}
First, we show the necessary information and then the interactions between the
involved entities.  The information and operations described in this section
apply to every data packet. \autoref{fig:acdcheader} shows the corresponding
\name data packet header.

\begin{table}[!t]
\caption{Summary of Symbols and Notation}
\scriptsize
\centering
%\scalebox{0.9}{
\begin{tabularx}{0.45\textwidth}{rX}
      \hline\\[-2pt]
	$\mathit{P[i]}$ & Policy packet information inscribed by AS$_i$ on the path.\\[2pt]
	$\mathit{CIR}$ & Committed Information Rate of the Token Bucket.\\[2pt]
	$\mathit{CBS}$ & Committed Burst Size of the Token Bucket.\\[2pt]
	$\mathit{\nam[i]}$ & \name header information inserted by AS$_i$.\\[2pt]
	$\mathit{PK_i^{+}}$/$\mathit{PK_i^{-}}$ & Public/private key pair of AS$_i$.\\[2pt]
	$K_{\mathit{SD}}$ & Shared key between source and destination.\\[2pt]
	$K_i$ & Local secret key of AS$_i$, for data-plane operations.\\[2pt]
	$\hat{K_i}$ & Long-term secret key of AS$_i$, for control-plane operations.\\[2pt]
	H($\cdot$) & A collision-resistant hash function, SHA-3.\\[2pt]
	MAC$_{K}$($\cdot$) & Message Authentication Code using key $K$.\\[2pt]
	Sign$_{i}$($\cdot$) & Signature of AS$_i$ with private key $\mathit{PK_i^{-}}$.\\[2pt]
	$T_m$ & Protest Time Margin.\\[2pt]
	$X|^{(m)}$ & The $m$ Most Significant Bits of $X$.\\[2pt]
	$X|_{(m)}$ & The $m$ Least Significant Bits of $X$.\\[2pt]
\end{tabularx}
\label{table:notation}
\end{table}

\begin{itemize}

\item Source Timestamp: an indication for the time when the packet
has left the source AS\@. It is a 16-bit value at the
granularity of 1 second. It suffices to capture durations up to 18 hours,
hence it constrains the possible values for the Protest Time Margin
$T_m$ (we have chosen $T_m=12$ hours).

\item Sequence Number: a 24-bit monotonically
increasing packet counter inserted by the source AS. The first packet
of a communication channel gets the value 0.

\item Integrity Check Value (MAC$_{K_{\mathit{SD}}}$): 
an 8-bit MAC over the payload-length field (in the network-layer header) and the other \name
related information inserted by the source AS (source timestamp and
sequence number). The purpose of the MAC is to signal on-path header modification; 
it is computed with the shared key $K_\mathit{SD}$. 
Although the MAC length is short, we do not use the MAC to provide
integrity guarantees per packet, but to signal misbehavior over an
aggregate of packets. In \autoref{sec:security_anal}, we quantify
the security implications of this idea.
The payload length is included in the computation of the ICV so that the destination stores
packet headers only, not the whole packets. 

\item Nonce fields: a 4-bit value inserted by each AS on the path. 
It functions as an indicator of having forwarded the packet and to enable
detection of replay attacks; the values are chosen uniformly at random.

\item MAC fields (MAC$_{K_{i}}$): a 4-bit MAC inserted by each AS on the path.
The input to the MAC is the information that must be integrity-protected to securely prove a 
sending-rate violation in the protest phase: the packet length in the network-layer header,
the source's timestamp and sequence number in the \name header, and the nonce field.
The local secret key $K_i$ used to compute the MAC is maintained by each AS
independently.  As described earlier, we use short MACs to signal misbehavior
over an aggregate of packets. We will show that even a 1-bit MAC can be used
for our purpose (\autoref{sec:security_anal}). 
If a subsequent entity changes any of the previous information in the packet,
the MAC verification will fail.

\item NextAS: an 8-bit pointer to the position in the \name
header where the next AS on the path will insert its information.
The pointer is initialized by the source and each transit AS modifies
it accordingly. The 8-bit field suffices for inter-domain paths up to
256 hops; the average AS-path length today is 3.9 hops (3.5
hops) for IPv4 (IPv6)~\cite{as_path}.

\end{itemize}

The sequence numbers, timestamps, and nonces are used to provide loose
replay detection at the AS-level granularity. Replay detection reveals such an attack
in the protest phase; the purpose is not to have the destination AS drop
replayed packets. 
The monotonically increasing values of the sequence numbers together with the
timestamp values are used to detect replay attacks. Multiple occurrences of a
sequence number for the same timestamp reveal the replay. Furthermore, the
clock deviation check at each AS hop prevents an attacker from storing and
replaying the packet at a later point in time. The random nonces inscribed by each AS provide
information about the adversary's position on the path. Nonces localize the
adversary to a portion of the path, depending on which nonce fields repeat and
which change per replayed packet.  Furthermore, the short MAC fields serve as a
misbehavior flag (rather than as integrity guarantees per packet): a few
verification failures in the protest phase indicate misbehavior.
Sections~\ref{sec:proof} and~\ref{sec:security_anal} provide further details.

\tinyskip\noindent
\textit{Processing of Outbound Packets:}
The source AS creates a \name packet header and fills in its information.
The new packet header is placed between the network and transport-layer headers and is created
with a sufficient length to accommodate the information of the
transit ASes; this ensures that the packet length does not increase en route.
The AS-level path is known to the source AS, and each transit AS overwrites 
1 byte of the header.
Based on the destination address in the packet header, the border
router of the source AS determines the shared key with the
destination AS, the current packet count for this communication
channel ($\mathit{seqno}$), and the output port to forward the packet to.
\autoref{alg:sending} summarizes the operations that the source 
performs.

\begin{algorithm}[t]
\caption{Procedure 1: Processing of Outbound Packets}
\label{alg:sending}
\footnotesize
\begin{algorithmic}[0]
\Procedure{Send}{$\mathit{pkt}, \mathit{pkt\_hdr}, \mathit{\nam}$}
	\LineComment{$\mathit{pkt}$ refers to the whole packet}
	\LineComment{$\mathit{pkt\_hdr}$ contains the network-layer packet header}
	\LineComment{$\mathit{\nam[~]}$ is the \name header that the source creates}
	\LineComment{$\mathit{cnt}$ the packet counter per communication channel }
	\State $\mathit(port, K_{\mathit{SD}}, cnt) \gets$ lookup($\mathit{pkt\_hdr}$)
	\State $\mathit{\nam[0].time} \gets$ time()$|_{(16)}$
	\State $\mathit{\nam[0].seqno} \gets \verb!++!cnt|_{(24)}$
	\State $\mathit{\nam[0].icv} \gets$ MAC$_{K_{\mathit{SD}}}$($\mathit{pkt\_hdr.payload\_len}$\par
		\hskip 70pt $||~\mathit{\nam[0].time}~||~\mathit{\nam[0].seqno}$)$|^{(8)}$
	\State $\nam.nextAS \gets 0$
	\State transmit($\mathit{pkt}, \mathit{port}$)
\EndProcedure
\end{algorithmic}
\end{algorithm}

\tinyskip\noindent
\textit{Processing of Forwarding Packets:}
We describe the actions that each egress border router of the
cooperating ASes on the path (AS\textsubscript{i}, $1 \leq i < n$)
performs for each data packet.

\begin{enumerate}

  \item Check the source's timestamp in the received
  packet and compare it with the local time. If the
  difference is greater than 3 seconds, drop the packet, otherwise
  forward the packet according to Step (ii). Step (i) ensures
  that the source is not indicating false timestamps.

  \item Add a short nonce (4 bits) and a MAC (4 bits)
  at the corresponding AS-specific position in the header.

  \item Increment the $\mathit{nextAS}$ pointer.

\end{enumerate}

Note that transit ASes do not need to perform
destination-based key switching since they use their local secret to
mark transit traffic. 
A non-cooperating AS ignores the \name header
and forwards the packet according to the destination address.
\autoref{alg:forwarding} summarizes these operations.

\begin{algorithm}[t]
\caption{Procedure 2: Processing of Forwarding Packets}
\label{alg:forwarding}
\footnotesize
\begin{algorithmic}[0]
\Procedure{Forward}{$\mathit{pkt}, \mathit{pkt\_hdr}, \mathit{\nam}$}
	\LineComment{$\mathit{pkt}$ refers to the whole packet}
	\LineComment{$\mathit{pkt\_hdr}$ contains the network-layer packet header}
	\LineComment{$\mathit{\nam[~]}$ is the \name header}
	\State $\mathit{diff} \gets |\mathit{\nam[0].time} - \text{time()}|_{(16)}|$
	\If {$\mathit{diff} > 3 \text{ and } \mathit{diff} < 2^{16} - 3$}
		\State drop packet
	\EndIf
	\State $\mathit{\nam[i].nonce} \gets$ rand()$|^{(4)}$
	\State $\verb!++!\mathit{\nam.nextAS}$
	\State $\mathit{\nam[i].mac} \gets$ MAC$_{K_i}(\mathit{pkt\_hdr.payload\_len}~||~\nam[0].time$ \par
	\hskip 70pt $||~\nam[0].seqno~||~\mathit{\nam[i].nonce})|^{(4)}$
	\State $\mathit{port} \gets$ lookup($\mathit{pkt\_hdr}$)
	\State trasmit($\mathit{pkt, port}$)
\EndProcedure
\end{algorithmic}
\end{algorithm}

\tinyskip\noindent
\textit{Processing of Inbound Packets:}
The destination performs the following data-plane operations:
\begin{enumerate}

  \item Check the timestamp, similar to transit
  ASes.

  \item Detect sending policy violations per established
  communication channel. This is straightforward by using Token
  Bucket as a policer, given the $\mathit{CIR}$ and $\mathit{CBS}$
  values. 

  \item Verify MAC$_{K_{\mathit{SD}}}$ to ensure that
  the source's information has not been modified en route. 
\end{enumerate}

The destination stores the packet headers (network-layer
headers and \name headers) as they contain potential proofs
of misbehavior. \autoref{alg:receiving} summarizes these steps.

\begin{algorithm}[t]
\caption{Procedure 3: Processing of Inbound Packets}
\label{alg:receiving}
\footnotesize
\begin{algorithmic}[0]
\Procedure{Receive}{$\mathit{pkt\_hdr}, \mathit{\nam}$}
	\LineComment{$\mathit{pkt\_hdr}$ contains the network-layer packet header}
	\LineComment{$\mathit{\nam[~]}$ is the \name header}

	\State $\mathit{diff} \gets |\mathit{\nam[0].time} - \text{time()}|_{(16)}|$
	\If {$\mathit{diff} > 3 \text{ and } \mathit{diff} < 2^{16} - 3$}
		\State drop packet
	\EndIf

	\State $icv \gets$ MAC$_{K_{\mathit{SD}}}$($\mathit{pkt\_hdr.payload\_len}$\par
		\hskip 70pt $||~\mathit{\nam[0].time}~||~\mathit{\nam[0].seqno})|^{(8)}$
	\If {$\mathit{icv} \neq \mathit{\nam[0].icv}$}
		\State drop packet
	\EndIf
\EndProcedure
\end{algorithmic}
\end{algorithm}

\sskip
\paragraph{2) Control Plane}
We present a policy setup that introduces no latency in the data
plane between the communicating end hosts of source and destination
ASes. The setup is based on two concepts. First, ASes advertise their
IP prefixes through BGP together with a default sending policy that
is used until source and destination ASes establish a new sending
policy with different properties. Second, using mostly symmetric-key
cryptography keeps the setup latency low. Specifically, only source
and destination ASes sign the sending policy with their private
keys, making the policy details provable and non-repudiable.
Transit ASes insert MACs that remind them of being on
the path of the communication channel. The combination of the
aforementioned concepts allows end hosts to communicate without
waiting for a sending policy setup and guarantees that the latency of
the setup remains low.

First, we summarize all the information that is required and then we show how
the policy is constructed.
\begin{itemize} 
	\item Current timestamp: inserted by the source AS,
	indicating the current time as the start for the communication
	channel.  
	\item Expiration timestamp: inserted by the destination AS,
	indicating the end of the communication channel.  
	\item Token Bucket properties: $\mathit{CIR}$ and $\mathit{CBS}$
	values are inserted by the destination AS and specify the sending
	properties for the source (see \autoref{sec:phase1}).
	\item \name-AS path: the source AS inserts the list of cooperating ASes
	on the path to the destination, which is known through the BGP
	advertisements.  
	\item Autonomous System Numbers: each AS on the path
	inserts its own ASN that serves as an identifier.  
	\item Signatures: source and destination ASes insert a signature over the policy details.
\end{itemize}
We provide more details about how this information is used.  

The source AS creates a policy packet ($P$) and sends it to the destination AS.
$\mathit{P[0]}$ corresponds to information inserted by the source AS and $\mathit{P[n]}$
to information inserted by the destination AS. 

\begin{enumerate}
	\item The source AS creates a policy packet $P$, with a timestamp indicating the start for the communication channel. 
	Moreover, the source inscribes its Autonomous System Number $\mathit{ASN_0}$, the current time, the cooperating ASes on the path, 
	and signs all the information with its private key $\mathit{PK_0^{-}}$.
	In particular, to avoid length-dependent security issues with signatures the hash of the information is signed~\cite{fdh_sec}.

	{\small
	\begin{align*}
		&\mathit{P[0].asn} \gets \mathit{ASN_0}\\
		&\mathit{P[0].time} \gets \text{time()} \\
		&\mathit{P[0].path} \gets \mathit{AS\_path}\\
		&\begin{multlined}
			\mathit{P[0].sig} \gets \sign{0}{\hash{\mathit{P[0].asn}\,||\,\mathit{P[0].time}
				\,||\,\mathit{P[0].path}}}
		\end{multlined}
	\end{align*}
	}

	\item Each transit AS\textsubscript{i}, $1 \leq i < n$, indicates its presence on the path. 
	It adds its $\mathit{ASN_i}$ and inserts a MAC over all the previous information.
	The MAC is computed with a long-term local secret ($\hat{K_i}$), known only to AS\textsubscript{i}, 
	that is used for control-plane operations.

	{\small
	\begin{align*}
		&\mathit{P[i].asn} \gets \mathit{ASN_i}\\
		&\begin{multlined}
			\mathit{P[i].mac} \gets \mathit{MAC}_{\hat{K_i}}(\hash{\mathit{P[0]}\,||\,\cdots\,||\,\mathit{P[i-1]}\,||\,\mathit{P[i].asn}})
		\end{multlined}
	\end{align*}
	}

  \item The destination AS receives $P$ and leverages the RPKI to verify
  the signature of the source AS. If verification succeeds, it
  fills in its $\mathit{ASN_n}$, the expiration time, and the Token
  Bucket values of $\mathit{CIR}$ and $\mathit{CBS}$. The destination
  signs the contents of the final sending policy and sends it back to the
  source AS.

	{\small
	\begin{align*}
		\centering
		&\mathit{P[n].asn} \gets \mathit{ASN_n} \\
		&\mathit{P[n].expiration} \gets \mathit{futureTime} \\
		&\mathit{P[n].CIR} \gets \mathit{CIR} \\
		&\mathit{P[n].CBS} \gets \mathit{CBS} \\
		&\begin{multlined}
			\mathit{P[n].sig} \gets \sign{n}{\hash{\mathit{P[0]}\,||\,\cdots\,||\,\mathit{P[n-1]}\,||\,\mathit{P[n].asn}\\
				||\,\mathit{P[n].expiration}\,||\,\mathit{P[n].CIR}\,||\,\mathit{P[n].CBS}}}
		\end{multlined}\\
		%&P[n].sig \gets \sign{n}{P[n].policy}
	\end{align*}
	}

	\item Source and destination ASes use the RPKI
	and perform a non-interactive Diffie-Hellman key exchange to derive a 
	shared key ($K_\mathit{SD}$) between them. 
\end{enumerate}

Note that transit ASes do not store information about the
sending policy, and only indicate their presence in the
communication channel. Moreover, only cooperating ASes indicate
their presence in the communication channel. The source AS stores the
final $P$ for at least a period of $T_m=12$ hours, as it is needed in
the protest phase.

The signatures and MACs, by which each entity authenticates the
information of all the previous entities, protect against path
falsification attempts. A malicious entity cannot substitute the
information inscribed by previous entities without invalidating
the signatures or MACs. To avoid malicious entities from truncating
on-path ASes, the source AS inserts the cooperating ASes on the path
($P[0].\mathit{path}$). In this way, on-path entities cannot truncate
on-path ASes, as the source has indicated which ASes will cooperate.
In addition, the source cannot lie and remove cooperating ASes from
the indicated path, as these ASes will inscribe their information and
reveal their support. Furthermore, the two timestamps indicate the
validity period of the channel so that complaints are temporally
confined.

\subsection{Verifying Proofs of Misbehavior}
\label{sec:proof}

In \autoref{subsec:phase3} we describe how the information in control and data
plane is used to prove misbehavior. In this section, we describe the operations
to examine a misbehavior report.

Recall that the information in the policy contains the transmission properties
($\mathit{CIR}$ and $\mathit{CBS}$) for the communication channel.
The data packet headers contain information for the actual
transmission properties. The transit ASes examine the received information as follows.

\begin{enumerate}
  \item ASes verify the signatures of the source and destination ASes
  in the policy packet, by obtaining the corresponding keys 
  from the RPKI.

  \item ASes verify the 4-bit MAC that they inscribed in the header.
  If all verifications succeed, ASes proceed with
  Step (iii). MAC verification failures signal en-route misbehavior 
  from a subsequent AS on the path from the source to the destination. In the 
  next section, we analyze scenarios with on-path malicious ASes.

  \item The ASes check conformance to the Token Bucket
  properties by running Token Bucket as a policer and
  by using the timestamp and payload length
  information of the headers.

\end{enumerate}

After the three-step procedure, the AS provides a signed admission or
rejection for the misbehavior to the reporting AS. The destination
AS collects the signed responses and sends them
back to all ASes on the communication channel.

\section{Protocol Analysis}
\label{sec:analysis}
This section analyzes the security and scalability properties of \name.
\subsection{Security Aspects} 
\label{sec:security_anal} 
We analyze the security properties of short MACs and then describe to
which extent \name is robust under two different threat models. We
first consider a strong threat model in which all entities can be
malicious. We then consider a second threat model that is slightly
weaker, but specifically designed to address current attacks.

\begin{itemize}

\item Threat Model I: Misbehavior is provable at least to the benign
cooperating ASes adjacent to the destination, under the strong threat model in
which source, transit, and destination ASes can be malicious and collude.

\item Threat Model II: Misbehavior is provable to \textit{all} cooperating
ASes on the path, under a weaker threat model in which transit ASes are not
malicious.

\end{itemize}

Our goal to present a deployable high-performance system deals with
the natural tradeoff between performance and security: some related
approaches provide stronger security guarantees, but come at the cost
of introducing considerable overhead. See \autoref{sec:related} for
the details on related work. 

\sskip
\paragraph{1) On the use of short MACs} 
Before discussing the two threat models in detail, we evaluate the
choice of short MACs.
Specifically, we argue that a very short MAC is sufficient to
provide accountability proofs in the context of flooding attacks.  There are
two important points to mention: i)~The role of the MACs in the packet, as
mentioned before, is only to provide a reminder to the transit ASes that they
have forwarded the packet. In the context of flooding attacks, we care about an
aggregate of packets and the collective proof that is constructed from this
aggregate, rather than from single packets. ii)~The secret keys used by other
ASes are unknown to the attacker. This means that an attacker can at best
randomly generate MACs without a means to check their validity.

The short length of the MACs does not prevent an attacker from generating valid
MACs. However, for an 8-bit MAC, 99\% of the generated MACs will not verify in
the protest phase and the misbehavior will thus be detected.

Taking this approach to the limit, we could use 1-bit MACs for
our purpose.  An attacker would have a 50\% probability to create
a valid MAC. Thus, 50\% of the crafted MACs would be invalid
(compared to 99\% previously) and the misbehavior is detected because of these invalid MACs. Our
choice of the MAC lengths is based on engineering the protocol
for high forwarding performance (byte aligned packet length), as
we show in \autoref{subsec:eval}.

\sskip
\paragraph{2) Threat Model I} 
We first analyze the scenario of colluding ASes and then two scenarios with
malicious transit ASes.

\begin{figure}[!t]
\centering
\includegraphics[width=.85\columnwidth]{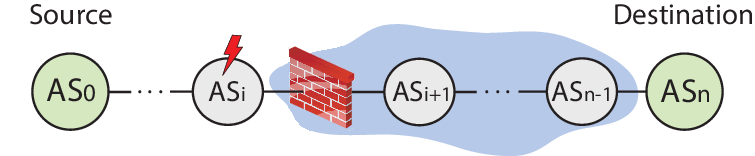}
\caption{\name operation with AS$_i$ being malicious.}
\label{fig:mal_networkmodel}
\end{figure}

\noindent
\textit{AS Collusion:} In this scenario, a transit AS colludes with a malicious
source AS to conceal an ongoing attack. The source is violating the sending
policy and transit AS$_i$ (\autoref{fig:mal_networkmodel}) corrupts all the
MACs of the previous ASes in the packet, causing verification failures when the
destination protests to these ASes. Hence, the policy violation cannot be
proven to these ASes. The first complaint round is successful only to the
shaded ASes in \autoref{fig:mal_networkmodel}, as AS$_i$ cannot corrupt MACs of
the subsequent ASes on the path. This limits the effectiveness of the proposal,
however, successful complaints even to a few transit ASes yield benefits, as
they can for instance install blocking filters closer to the source, as
depicted in \autoref{fig:mal_networkmodel}.

The above scenario presents the worst case, in which a transit AS corrupts all
previous MACs.  If AS\textsubscript{i} does not corrupt all the previous MACs, complaining
is more effective since more ASes would acknowledge the attack. 
\textit{Notice that the complaint is accepted at least by the
benign cooperating ASes adjacent to the destination.}
Hence, collusion with multiple ASes does not provide additional benefits
to the source, as the effectiveness of the proposal depends only on the
position of the malicious AS that is closer to the destination. 

\tinyskip\noindent
\textit{Packet Replay:} In this scenario, we assume that a malicious
transit AS forwards a packet multiple times to increase
traffic and thus to blame an innocent source AS.

A packet replay is indicated through multiple occurrences of the same sequence
numbers for a given timestamp. Furthermore, the clock deviation check does not
allow an adversary to store packets and replay at a later time.  The 24-bit
sequence number suffices for more than $16\cdot 10^6$ packets and the monotonically
increasing values render multiple occurrences per timestamp suspicious. For
example, a communication channel with a $\mathit{CIR}$ value of 1~Gbps has an
average packet-sending rate of 325~kpps for the average packet length of 413
bytes~\cite{caida}. For an attack where each packet is replayed twice, there
are on average $10^6$ packets that belong to each slot of 1 second, but the 24-bit
field suffices for more than $16\cdot 10^6$ packets. Under normal operation each sequence
number would show up only once, but multiple occurrences indicate a replay. 

A high-sending-rate policy that uses up the
available nonce space in the slot of 3 seconds would possibly
allow an attacker to replay packets, but this is very unlikely:
For the average packet size of 413 bytes it would require a
communication channel of 17~Gbps for this to happen, which is an
unrealistic value for a single channel. If such throughput values
become reality in the future, increasing the sequence number length
will solve the problem. For instance, 32 bits suffice for over 4 billion
packets.

The nonce fields are used for detecting the adversary's
location on the path: If AS$_i$ (\autoref{fig:mal_networkmodel})
replays packets, then the combinations of sequence number and nonce
field of only the first $i-1$ ASes occur multiple times. In other words,
the location of the attacker can only be between AS\textsubscript{i} and
AS\textsubscript{i+1}.
The reason is twofold. First, non-cooperating ASes between
AS\textsubscript{i} and AS\textsubscript{i+1} might replay packets.
Second, the attacker (AS\textsubscript{i}) might inscribe nonces in a
way that puts the blame on the next AS (AS\textsubscript{i+1}).
Hence, the localization cannot identify the attack to a specific
entity, but all ASes after the replaying AS become aware of the
approximate location of the attack and can take action.

Note that we use sequence numbers and nonce fields to detect replay
attacks in the protest phase, rather than to drop replayed data-plane traffic.

\tinyskip\noindent
\textit{Packet Injection:} In this attack, a transit
AS attempts to craft fraudulent packets and
inject them into the network. This attack is prevented thanks to the
MACs inserted by the source and the transit ASes. Assuming that the
adversary has not obtained the local secrets of the other entities,
its probability of inserting only valid MACs is negligible, as
discussed before. The verification failures of inserted invalid MACs
will reveal the attack.

If AS\textsubscript{i} (\autoref{fig:mal_networkmodel}) injects traffic, the
subsequent ASes on the path insert their MACs as usual. These MACs will verify 
in the protest phase and hence the shaded ASes in \autoref{fig:mal_networkmodel}
acknowledge the violation, exactly as in the packet replay attack.

\sskip
\paragraph{3) Threat Model II} 
Attacks usually originate from malicious or vulnerable end hosts
inside the source AS; transit ASes usually have no incentive to
collude with other ASes, nor to engage in malicious conduct, such as
packet replay. The forwarding proof thus remains intact during
transit and \textit{all cooperating ASes on the path from the source to 
the destination acknowledge the attacks}.

\paragraph{Other Attacks} Here we describe some protocol
manipulation attacks that are specific to \name. 
Since the destination uses the received packets as a proof of an
attack, the source can craft timestamps in the packet, which
together with the aggregate traffic size do not violate the
policy. The clock deviation check protects against this, but
allows the source to shift timestamps by one second, only once
though. More specifically, the source can send excessive traffic
in the slot of one second by putting the timestamp of the next
second in some packets. In this way the maximum burst size
violation for one time interval is not detected, but it restricts
the traffic for the subsequent intervals, as it must be lower to
conform to the policy's $\mathit{CBS}$ value for the next
intervals.  A sending rate that exceeds the $\mathit{CIR}$ value
over any multiple of the time interval cannot be concealed. The
Token Bucket properties in combination with the clock deviation
check also protect from a coward attack~\cite{coward}; in a
coward attack the attacker scales down the intensity temporarily
to avoid detection.

Another general attack against accountability
frameworks consists in falsely blaming benign entities. A malicious
destination can try to convince transit ASes by providing multiple
times the same packets as evidence of increased traffic. This is a
variation of a replay attack and the sequence number and nonce fields
prevent it. Crafting the timestamps will cause MAC verification
failures and the transit ASes will not acknowledge the proof.

\subsection{Scalability} \label{subsec:overhead} 
We examine the scalability properties of \name in terms
of bandwidth and storage overhead.
Concerning the processing overhead, we provide a detailed evaluation in
\autoref{sec:impl}.

\sskip
\paragraph{1) Bandwidth Overhead} Our proposal comes at the cost of
increased packet size. The source AS inscribes a constant
amount of 7 bytes/packet and each transit AS adds another 1 byte. We
envision a \name integration with the IP protocol
and this would require two additional bytes per
packet only in the case of IPv6 traffic (more details, also on IPv4, follow in
\autoref{sec:impl}). To put this overhead
into context, we analyze three 1-hour packet traces of OC-192
backbone links obtained from CAIDA~\cite{caida}. We take a
pessimistic approach on the AS-path length to quantify the overhead
and assume it to be 5~hops.\footnote{RIPE Labs report an
average length of 3.9 hops for IPv4 and 3.5 hops for
IPv6~\cite{as_path}.} Based on the number of packets in IPv4
and IPv6 and their ratio on the link, we calculate the
link's overall bandwidth overhead.
\autoref{table:packetrace} shows the properties of the traffic on
the link and the overall overhead: the bandwidth overhead does not
exceed 2\%. This estimation assumes that the AS-path length is
independent of the packet length distribution.

\begin{table}[!t]
  \footnotesize
  \centering
  \scalebox{1}{
    \begin{tabular}{l|c|c|c}
      & Trace 1 & Trace 2 & Trace 3 \\ \hline\hline
      Trace rate~(Gbps)\! & 1.63 & 3.72 & 3.57 \\
      IPv4 pkt.~(bytes)&\!\!\! 747 (99.95\%) & 920 (99.96\%) & 736 (99.88\%)\! \\
      IPv6 pkt.~(bytes) & 130 (0.05\%) & 342 (0.04\%) & 155 (0.12\%) \\
      \textbf{BW overhead} & \textbf{1.71\%} & \textbf{1.39\%} & \textbf{1.74\%}
    \end{tabular}
  }
  \caption{Bandwidth overhead of \name for three backbone-link traces. 
  The reported sizes are mean values and the parentheses show
  the percentage of traffic for each IP version.}
  \label{table:packetrace}
  \vspace{-8mm}
\end{table}

\sskip
\paragraph{2) Storage Overhead} To provide a scalable framework, our
goal is to reduce the amount of state stored at the forwarding
devices of cooperating ASes. Source and transit ASes do not need to
store data-plane related information. The source stores one policy
packet and a shared key $K_\mathit{SD}$ (16 bytes) per
communication channel. The total number of ASes in the Internet is
less than 50,000~\cite{as_numbers}, which means minimal overhead
(800~kB) even if there is a communication channel with every other AS.

Furthermore, the transit ASes store only local secret keys
(independent of any communication channel). As noted in
\autoref{sec:security_anal}, there is no strict requirement on the
frequency of changing keys, however, the previous keys are kept
to verify MACs that were computed earlier.
According to the protocol, a cooperating AS accepts and examines
incoming proofs up to a period of $T_m=12$ hours in the
past. Hence, the storage overhead depends on the frequency with which the AS
changes its keys within the $12$-hour frame. For example, a transit AS that changes
its local keys ($K_i$, $\hat{K_i}$) every minute requires a storage
capacity of 250~kB for the $12$-hour period.

The most significant storage overhead occurs for the
destination AS when storing data packet headers as a proof
of source misbehavior. The destination can provide the proof to the
transit ASes up to 12 hours after it received the packets.  
For a destination AS that stores the IP and
\name packet headers of the 1-hour link traces in
\autoref{table:packetrace}, the storage requirement is 30.2, 56, and
67.3~GB, respectively. For this calculation, we assume again an
AS-path length of 5 hops and took into consideration the different
overhead of the IPv6 header (40~bytes) and the IPv4 header
(20~bytes), on top of the \name header overhead. 

Note that the considerable storage overhead is shifted to the destination AS
since it is in the destination's interest to be protected from flooding
attacks; thus having forwarding ASes store the packets would distribute the
storage overhead in an unfair manner.
Moreover, to further decrease the 
overhead, destination ASes store only packet headers. Also, the
destination can choose when to protest about a violation, hence it
does not have to store headers for 12 hours and can regulate the
storage requirement. In addition, ASes can store compressed proofs of
misbehavior only for the violated time periods instead of storing the
whole set of packets of the communication channel.

\section{Implementation and Evaluation}
\label{sec:impl}
We describe our protocol in the context of today's Internet, implement a 
software switch prototype, and evaluate performance on a server and
a desktop machine. 

\subsection{Integration with IP}
We analyze the deployment of \name with IP.  
IPv6 allows a straightforward and elegant implementation by using
Extension Headers (EHs)~\cite{rfc2460}. 
IPv6 Extension Headers encode optional IP-layer information in headers that are
placed after the regular IPv6 header.  They make the protocol extensible by
allowing support for security, mobility, and other services.

\begin{figure}[!t]
	\centering
	\includegraphics[width=\columnwidth]{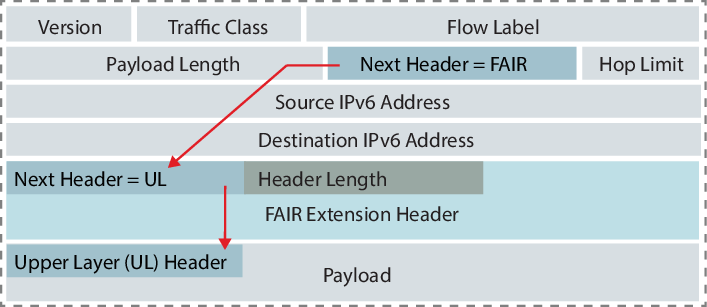}
	\caption{IPv6 packet with \name Extension Header.}
	\label{fig:acdc_ext}
\end{figure}

The IPv6 specification~\cite{rfc2460} defines some default EHs for additional
network-layer services and leaves space for new EHs. To implement
\name, we define a new EH that is processed only by egress border
routers of cooperating ASes. According to the specification, the
Hop-by-Hop EH is the only EH that \textit{must} be processed by all network
devices, whereas other EHs are inspected only by devices configured
for certain services. This feature allows ISPs to adopt \name in an
incrementally deployable fashion without breaking
legacy IPv6 traffic. \autoref{fig:acdc_ext} shows a regular IPv6
header together with the \name\ extension. The \name EH is placed after the
regular IPv6 header or after the Hop-by-Hop EH (if present), as the
IPv6 specification commands. The \texttt{Next Header} field (whether in the regular
header or in a preceding EH) points to the start of the \name EH.
The content of our EH is what \autoref{sec:protocol} describes and
\autoref{fig:acdcheader} depicts.  To make \name compatible with
IPv6, two additional fields are required: a pointer (8~bits) that
points to the next EH or to an Upper Layer (UL) protocol, and a \texttt{Header
Length} field (8~bits) that indicates the length of the EH\@. This
translates into an additional overhead of 2 bytes.

Extension Headers are considered an intrinsic part of IPv6 and the way they are
processed by network devices can harm forwarding performance.  However, IPv6
provides an elegant deployment path due to EHs. This feature is not supported
by IPv4 and a workaround for IPv4 is necessary.

IPv4 has inherent limitations with regard to extensibility which
complicates deployment. The \name header can be implemented as a
``shim'' layer between the IPv4 header and the transport protocol.
The border routers of source ASes insert the \name header after the
IPv4 header; border routers of transit ASes locate and process the
\name header, as it starts 20 bytes after the IPv4 header; and the
border routers of destination ASes store and remove the \name header
before forwarding the packet to the destination host. Shim-layer
approaches typically cause problems due to middleboxes in the source
and/or destination ASes~\cite{tracebox}. However, note that the \name
header is not visible inside those domains, alleviating such concerns.

\subsection{Software Switch Prototype}
To test the practicality of our proposal, we implement the required
functionality in software.  We recognize a resurgence of interest in software
switches thanks to their flexibility and programmability at low procurement and
operational costs~\cite{cuckoo,packetshader,routebricks}.  Furthermore, recent
advances in the software-switching field demonstrate that these advantages do
not come at the cost of performance, which has traditionally been the Achilles'
heel of software switches.  We use the Intel Data Plane Development Kit
(DPDK)~\cite{dpdk} as the packet I/O engine and take advantage of the Intel
AES-NI~\cite{aes_ni}.

The \textbf{Intel DPDK} is a high-performance packet I/O engine that provides
flexibility and programmability, allowing packet processing in user space.
DPDK uses polling to avoid the overhead of unnecessary interrupts. It provides
optimized Network Interface Card (NIC) drivers that map packet buffers directly
in user space to avoid redundant memory accesses (zero copy). We choose DPDK
for our development platform as it efficiently performs packet I/O and allows
us to focus on the \name EH processing.

The \textbf{Intel AES-NI} is a recent instruction set that uses hardware
support to speed up encryption and decryption of AES operations. Intel reports
a performance of 2.01 Cycles Per Byte (CPB) for a 16-byte block AES encryption
on an Intel Westmere running at 2.67~GHz~\cite{aes_ni}.

We describe the implementation of the necessary components for \name.
To construct the required MACs, we use the Cipher Block Chaining mode
(CBC-MAC) with AES as the underlying block cipher. 
The CBC-MAC encryption of a plaintext block depends on the encryption of the
previous block; the output is the final block. The value for the Initialization
Vector (IV) is 0. The size of both input blocks and the output block is 128
bits (16 bytes).
The input length to the CBC-MAC is fixed and independent of the AS path
length\footnote{CBC-MAC is insecure for variable-length
messages~\cite{cbc_sec}.}. Also, the input fits in one block (less than 16
bytes). Furthermore, the input length of the MACs in the control plane is fixed
as well. We use 128-bit encryption keys and keep only the required number of bits from
the output, as specified in \autoref{sec:protocol}.

The source AS of the outgoing traffic has to look up the shared key
with the destination ($K_{\mathit{SD}}$) and the current packet count for the
communication channel, as it is used for the sequence number
($\mathit{seqno}$). The source uses the shared key with the
destination in order to compute the MAC. To implement these
functionalities at line rate, we extend the Forwarding Information Base
(FIB) to contain not only the egress interface, but also the shared
symmetric key with the destination and the current value for the
sequence number. 

This increases the size of the FIB, but it still fits
in today’s SRAM caches, avoiding access to the substantially slower
DRAM. The size of the extended FIB for today's IPv4 BGP routing table
sizes is around 12~MB~\cite{ripefib} and for IPv6
around 1~MB~\cite{ripefib}, which is lower than SRAM sizes even on commodity
hardware, as we show in our evaluation. In addition, the increase in
length for each FIB entry does not degrade forwarding performance
since each FIB entry fits into the typical cache line of 64 bytes.
Even in case of IPv6 addresses, where each entry requires 36~bytes
(16-byte destination address, 16-byte symmetric key, 3-byte sequence
number, and 1-byte output interface). 

To generate randomness for the nonce and to mark fields at
line rate, we need an efficient pseudorandom number generator
(PRNG). We implement a thread-safe, multicore version of the Linear
Congruential Generator (LCG) that meets our performance requirements.
Modern CPUs come with Digital RNG (DRNG) hardware
implementations~\cite{drng} that can speed up this process
significantly~\cite{drngimpact}. Unfortunately, our CPUs lack
this feature.
Furthermore, each CPU core has an AES hardware unit. We assign each
core to handle one port, taking advantage of the processing power of
today's multicore systems.  For the timestamp, we use the least
significant bits (LSB) of the Unix time.

We bring these components together on two different machines: a commodity
server and a low-end desktop.  The server has a non-uniform memory access
(NUMA) architecture with two Intel Xeon E5-2680 CPUs that communicate over two
QPI links. Moreover, each NUMA node is equipped with four banks of 16~GB DDR3
RAM. In total, we have 6 dual-port 10~GbE NICs (PCIe Gen2 x8) that can provide
a maximum capacity of 120~Gbps. The total cost of this setup is around
\$$7,000$. \autoref{table:server_hwspecs} summarizes the hardware specification
of the server machine.

\begin{table}[t]
\small
\centering
\scalebox{0.9}{
\begin{tabular}{l|l|c|c}
      \textbf{Item} & \textbf{Model Name} & \textbf{Qty} & \textbf{Unit price} \\
      \hline
      \hline
      Board & Intel S2600GZ (2 sockets) & 1 & \$670 \\
      CPU & Intel Xeon E5-2680 (8 cores, 2.7~GHz) & 2 & \$1,727 \\
      RAM & Kingston DDR3 4~GB (1,333 MHz) & 8 & \$38 \\
      NICs & Intel 82599EB X520-DA2 10~GbE & 6 & \$450
\end{tabular}
}
\caption{Specification of utilized Server Hardware.}
\label{table:server_hwspecs}
\end{table}

The desktop machine is a Lenovo ThinkCentre Edge 3494AZG with an
Intel Core i5-3470S CPU  with one dual-port 10~GbE NIC (PCIe Gen2 x8)
and a total cost of \$$1,200$. 
\autoref{table:desktop_hwspecs} shows the hardware specification of
the desktop machine.

\subsection{Switch Prototype Evaluation}
\label{subsec:eval}
We evaluate the switching performance of both machines and demonstrate
that the EH processing incurs minimal computational overhead
even for low-end hardware.

In the experiments, we emulate traffic flows originated by a source AS
and evaluate the performance of a \name-enabled border router. 
We evaluate the worst case, and thus we use IPv6 that is slower than IPv4
because the Forwarding Information Base (FIB) entry is longer than for IPv4;
we have observed the same forwarding performance also for IPv4 traffic.
Moreover, we specify random destination addresses for the generated flows,
eliminating spatiotemporal locality for cache accesses. 
Using random destination addresses captures any performance degradation due to
key switching with different destination ASes. 
To generate traffic, we use Spirent SPT-N4U-220 as our packet
generator. The table lookup is performed by an implementation of
DIR-24-8-BASIC~\cite{dir24} for IPv6 addresses. We generate the FIB from 
a BGP routing table snapshot (November 2014) from RIPE RIS, with 18k
unique IPv6 prefixes~\cite{ripefib}.

\begin{table}[t]
\small
\centering
\scalebox{0.9}{
\begin{tabular}{l|l|c|c}
      \textbf{Item} & \textbf{Model Name} & \textbf{Qty} & \textbf{Unit price} \\
      \hline
      \hline
      CPU & Intel Core i5-3470S (4 cores, 2.9~GHz)& 1 & \$170 \\
      RAM & Hynix DDR3 4~GB (1,600~MHz) & 1 & \$45 \\
      NICs & Intel 82599EB X520-DA2 10~GbE & 1 & \$450
\end{tabular}
}
\caption{Specification of utilized Desktop Hardware.}
\label{table:desktop_hwspecs}
\end{table}

First, we evaluate the performance of a single 10G port for
three packet sizes; then we enable all ports. Finally, we evaluate
performance with all ports enabled and for varying packet sizes. All the 
experiments are conducted on the server and the desktop platforms.

\begin{figure}[!t]
	\centering
	\includegraphics[width=\columnwidth]{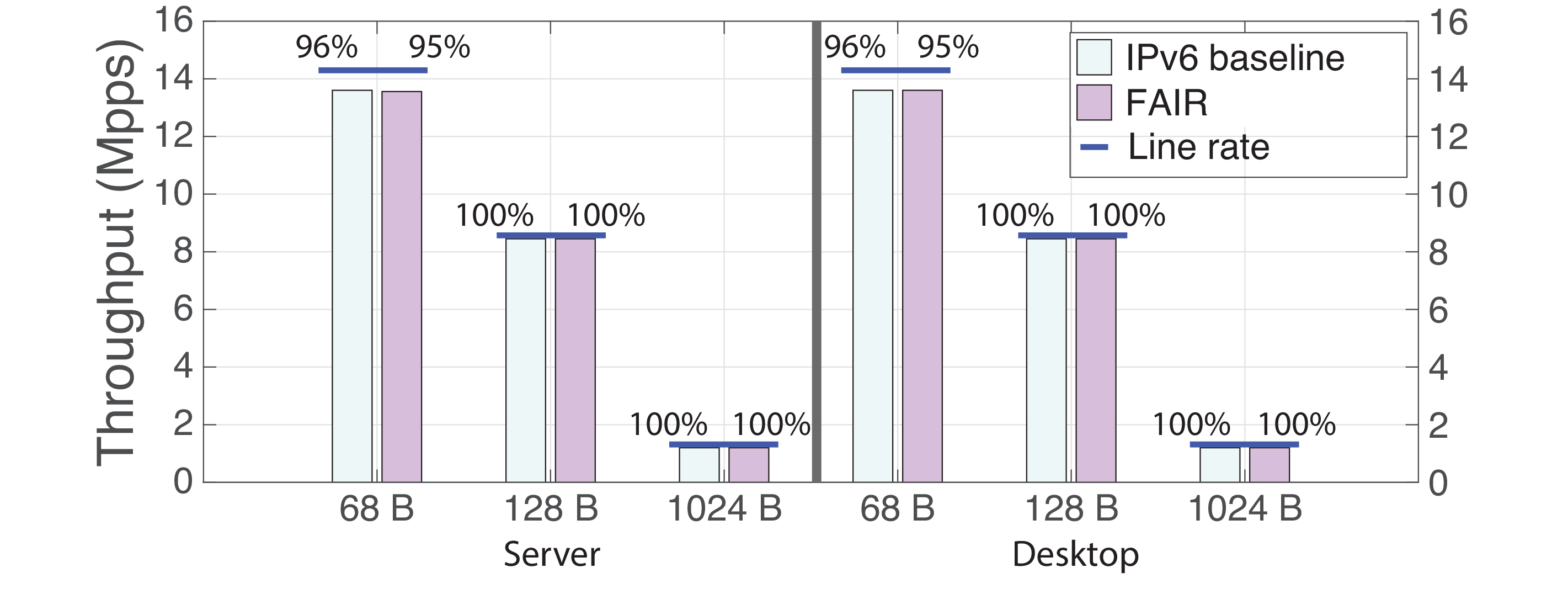}
	\caption{Switching performance of the server and the desktop for one port activated, and 68, 128, and 1024-byte packets.}
	\label{fig:port1}
\end{figure}

\sskip
\paragraph{1) Single-port experiment} 
First, we test the switching performance of one port for three packet sizes:
68, 128, and 1024 bytes. Minimum-sized packets, 68 bytes, translate to a higher packet rate and
are the worst case for the EH processing.  The minimum length for IPv6
packets with the \name EH is 68 bytes (instead of 64) due to the
additional information. \autoref{fig:port1} shows the switching performance
for the server and the desktop platform.

The highest packet rates for the three packet sizes are 14.20~Mpps,
8.45~Mpps, and 1.20~Mpps on a 10~GbE link; we refer to these values as the line-rate performance. 
The baseline for the experiments is the
switching performance of legacy IPv6 traffic (only table lookup and
forwarding). The figure shows that the EH processing degrades
performance by only 1\% for minimum-sized packets on both machines. The figure also shows the line-rate
performance (blue line) and the minimal baseline degradation due to
the table lookup and the high packet rate for the 68-byte case. For the longer
packet sizes, the switching performance reaches the line rate on both machines.
The single-port experiment demonstrates that switching performance is close to optimal
for one port, even on low-end hardware. Next, we increase the switching load.

\begin{figure}[!t]
	\centering
	\includegraphics[width=\columnwidth]{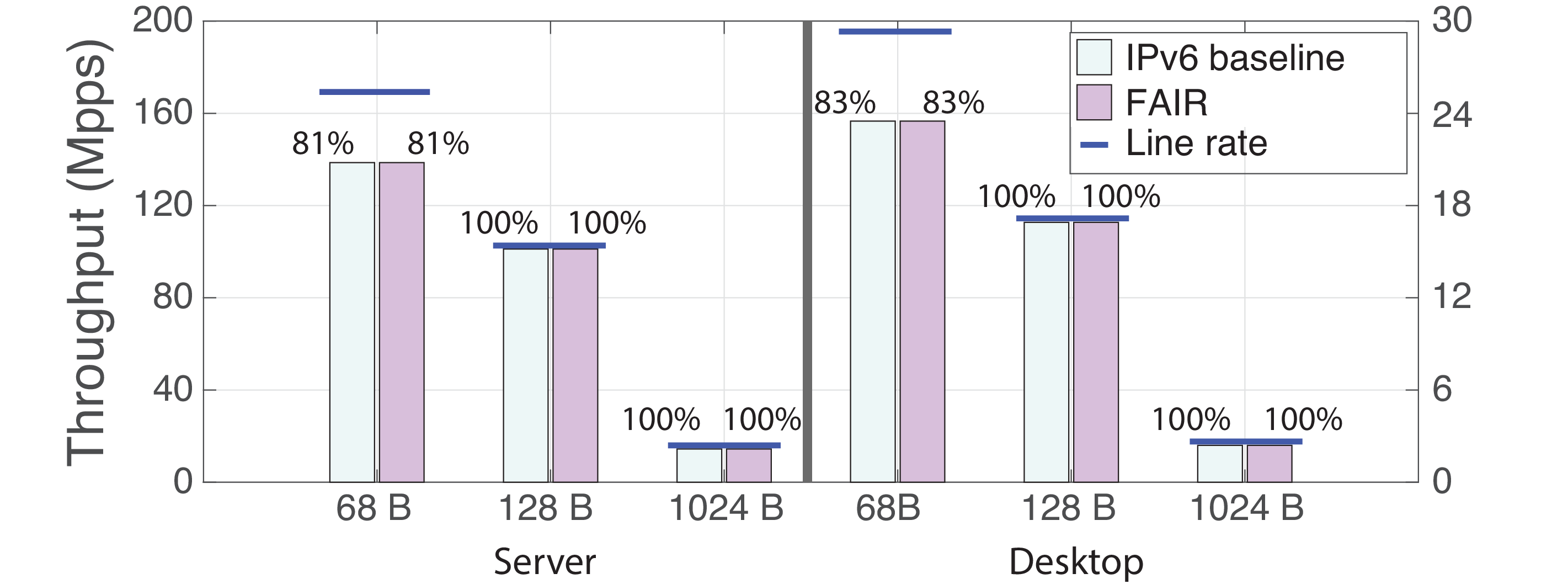}
	\caption{Switching performance of the server and the desktop for all ports activated, and 68, 128, and 1024-byte packets.}
	\label{fig:port12}
\end{figure}

\sskip
\paragraph{2) All-ports experiment}
To demonstrate that the \name EH processing scales for increasing packet
rates, we activate all ports; each port is served by a different
CPU core. Again we use the same three packet sizes.  
\autoref{fig:port12} shows the results.

We use a different scale in the figure for the two machines, since they accommodate a different number of ports.
The packet line rates for the server (12 ports) and the three packet sizes
are 170.4~Mpps, 101.4~Mpps, and 14.4~Mpps, respectively. The packet line rates for the
desktop (2 ports) and the three packet sizes are 28.40~Mpps, 16.90~Mpps, and 2.40~Mpps, respectively.
We see that throughput scales for multiple ports and \name switches at baseline performance
for the three packet sizes, on both machines. The experiment demonstrates how switching performance
scales for increasing packet rates, even for the low-end hardware.
However, we notice a higher baseline degradation for 68-byte packets: in the one-port experiment,
the switching performance was at 96\% of the line rate, whereas now it
is around 80\%. The explanation is that our I/O subsystem hits a
bottleneck when both ports of a NIC receive packets at the maximum
packet rate. The bottleneck exists irrespective of \name: the PCIe Gen2 x8 interface of our
NICs cannot sustain this packet rate when both ports are active.  The
packet rate of each port is capped at 11.55~Mpps.
Cuckooswitch~\cite{cuckoo} uses the same NICs and reports the same
limitation.

\sskip
\paragraph{3) CPU as the bottleneck}
To bypass the I/O bottleneck and stress the limits of the CPU, we assign the traffic from
two ports of different NICs to one core; this makes the CPU the throughput bottleneck.
For minimum-sized packets, the CPU handles 21.62~Mpps out of the maximum 28.40~Mpps. 
Hence, one CPU core can process traffic from more than one 10~GbE port that
receives packets at the maximum packet rate.

\begin{figure}[tp]
	\centering
	\includegraphics[width=.75\columnwidth]{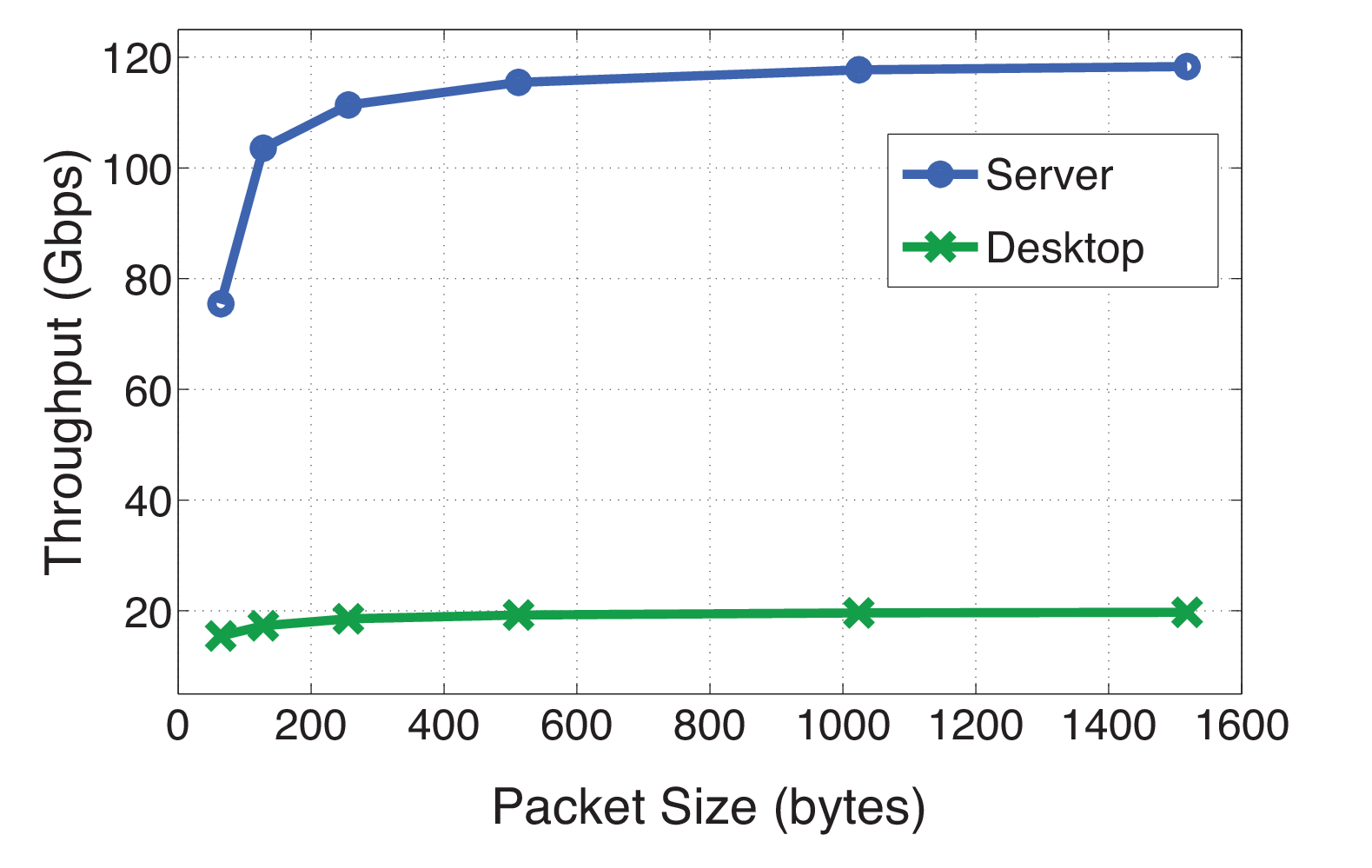}
	\vspace{-2mm}
	\caption{Switching performance for all ports.}
	\label{fig:gbpsall}
	\vspace{-4mm}
\end{figure}

Next, we show that for increasing packet sizes, \name saturates
line-rate bandwidth and achieves 120~Gbps and 20~Gbps for the server
and desktop respectively. \autoref{fig:gbpsall} shows the
throughput for 68, 128, 256, 512, 1024, and 1518-byte packets. We omit
the line-rate line; for all measurements --- except the 68~byte packet
--- it is identical to the drawn lines. Hence, as we increase the packet size and
the packet rate drops, IPv6 baseline and \name performance is at 100\% line rate.

\section{Protection from DDoS Attacks}
\label{sec:bit}

\name, as an accountability framework, does not provide active protection
from attacks, as it does not enforce specific behavior when an attack
is detected. This section describes a more radical application of \name that enforces
and pushes higher security standards to the edge of the Internet.
Furthermore, the section illustrates how \name can be combined with active
defense mechanisms.

\subsection{Suspicious Bit}

The April Fool's proposal of the ``evil bit''~\cite{rfc3514} describes a
security mechanism from an idealist's point of view: data packets carry a
security flag -- the \emph{evil} bit -- to indicate malicious intent; the
flag is set by the malicious senders themselves. 

We propose a more realistic security mechanism, the \textit{suspicious}
bit that is set by transit ASes to indicate suspicious traffic. With 
such a mechanism in place, the traffic itself becomes the indicator of 
possibly malicious behavior and incentivizes transit ASes to take action.
For instance, an AS can drop or deprioritize suspicious traffic in case
of congestion, ensuring better service for its benign customers. In addition,
flagging traffic due to an attack on one victim provides protection to other
potential victims as well.

An immediate question is how ISPs distinguish
benign from suspicious ASes in order to flag their traffic.
We leverage \name as a building block to address this question.
\name's initial sending policy negotiation provides
a clear line for detection of misbehavior;  
the \name header in the data
packets provides the corresponding accountable proofs of misbehavior.

Another question is how ISPs are incentivized to flag their misbehaving customers.
The answer to this question lies in the competitive environment in
the Internet ecosystem. 
Recall that \name's accountable proof of misbehavior is received by
all on-path ASes. If an ISP does not flag its provably malicious
transit traffic, then the next AS on the path will
flag \textit{all} of the traffic of the previous AS. We believe that the threat of collateral
damage and the harsh competitive Internet market pushes ISPs to mark their
customers' traffic. If innocent customers experience packet drop because of
their ISPs' poor security practices, they have an incentive to switch to a
more reliable ISP, if possible.

\begin{figure}[tp]
        \centering
        \includegraphics[width=.60\columnwidth]{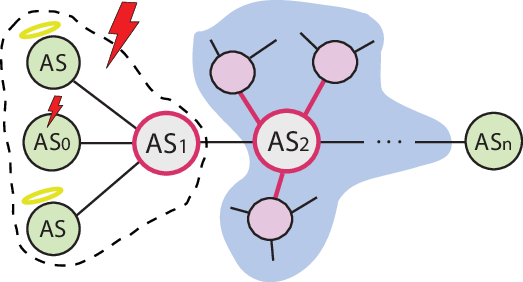}
	\caption{The suspicious bit identifies traffic from a portion of the network with poor security practices.}
	\label{fig:suspicious}
\end{figure}
We emphasize that ISPs do not have to drop suspicious traffic right away for
two reasons.  First, the suspicious bit indicates only that the traffic is
suspicious (not necessarily malicious) and thus gives incentives to take action under certain conditions
(e.g., drop it in case of congestion). Second, under the strong threat model, an
adversary could set the bit for legitimate traffic to make another ISP drop the traffic.
Consequently, setting the suspicious bit for legitimate traffic would not be a
useful attack strategy. In addition, today's Internet is opaque to loss
anyway~\cite{audit}, and hence the adversary can directly drop the traffic
and evade detection.

We demonstrate the suspicious bit application by means of
\autoref{fig:suspicious}. The illustrated network topology shows
malicious AS$_0$ violating the sending policy negotiated with benign
AS$_n$. AS$_1$ is the ISP of the malicious AS$_0$ and other benign
ASes. It hence provides transit to more than a single customer.
Assume that AS$_1$ has received a proof of misbehavior for AS$_0$: 
AS$_n$ has reported malicious traffic to AS$_1$. In the
ideal case, AS$_1$ would mark traffic from AS$_0$ as suspicious,
warning other entities in the network. If, however, AS$_1$ does not
mark the suspicious traffic, then AS$_2$ will mark \emph{all} the
traffic from AS$_1$ as suspicious.

This overstatement, however, means that also traffic from the \emph{benign} customers
of AS$_1$ gets flagged as suspicious, which will lead to collateral damage if
a downstream ISP decides to drop traffic. By flagging traffic, AS$_2$ informs other
entities in the network (shaded part) that some portion of the network
(dashed part) might be misbehaving.  This practice will incentivize AS$_1$ to behave
correctly and to flag the traffic of its misbehaving clients, thereby protecting
its benign clients.
As a consequence, the stub ASes are pushed to deal with
their internal security issues (e.g., botnets inside an AS or
misconfigured services) to protect the innocent flows from being dropped.

\subsubsection*{Forwarding with the Suspicious Bit}

We show the information and data structures when forwarding traffic
under the SB application.

\begin{itemize}

  \item Suspicious Bit ($\mathit{sb}$): the SB flag, used to mark a
  packet as suspicious, is the most significant bit of the
  $\mathit{nextAS}$ pointer in the \name header. This means that
  routers will check and update the 7 least significant
  bits of the pointer, which suffice to encode AS-paths
  of length up to 128 hops.

  \item Suspicious Sources ($\mathit{sus\_sources}$): set of
  addresses for which the AS has acknowledged the violation.

  \item Suspicious Ports ($\mathit{sus\_ports}$): set of the switch's
  ports that receive traffic from an insecure part of the network. We refer
  remaining ports of the switch as non-suspicious.
\end{itemize}

In the following, we describe how this information is used to realize the
suspicious bit application. Note that the SB does not enforce a specific
action, hence the transit AS can forward, drop, or delay traffic based on
its traffic engineering and security policies. 
\autoref{alg:sbalg} provides the pseudocode for traffic forwarding with the suspicious bit.

\begin{itemize}

	\item If incoming traffic arrives at a non-suspicious port:
		\begin{itemize}
			\item if the SB is set then forward/drop/delay traffic.
			\item if the SB is not set and the source address belongs to the suspicious sources then
			add the port to the suspicious ports. Set the SB and forward/drop/delay traffic.
		\end{itemize}
	\item If incoming traffic arrives at a suspicious port:
		\begin{itemize}
			\item if the SB is set, remove the incoming port from the suspicious ports. In this way
			if previous ASes that did not flag traffic start
flagging, their whole traffic is not flagged 
			as suspicious anymore. Then forward the traffic.
			\item if the SB is not set, then set the SB. Then forward/drop/delay.
		\end{itemize}	
\end{itemize}

\begin{algorithm}
\caption{Procedure 4: Forwarding packets in the SB application}
\label{alg:sbalg}
\footnotesize
\begin{algorithmic}[0]
\Procedure{Forward}{$\mathit{pkt\_hdr}, \mathit{\nam}, \mathit{port\_in}$}
	\LineComment{$\mathit{pkt\_hdr}$ contains the network-layer packet header}
	\LineComment{$\mathit{\nam}$ is the \name header}
	\LineComment{$\mathit{port\_in}$ is the ingress port of the packet}

	\If {$\mathit{port\_in} \not \in \mathit{sus\_ports}$}
		\If {$\mathit{\nam.sb}$}
			forward/drop/delay traffic
		\Else
			\If {$\mathit{pkt\_hdr.src\_addr} \in \mathit{sus\_sources}$}
				\State $\mathit{sus\_ports} \gets \mathit{sus\_ports} \cup \{port\_in\}$
				\State $\mathit{\nam.sb} \gets 1$
				\State forward/drop/delay traffic
			\Else
				\State forward traffic
			\EndIf
		\EndIf
	\Else
		\If {$\mathit{\nam.sb}$} 
			\State $\mathit{sus\_ports} \gets \mathit{sus\_ports} - \{port\_in\}$
			\State forward traffic
		\Else
			\State $\mathit{\nam.sb} \gets 1$	
			\State forward/drop/delay traffic
		\EndIf
	\EndIf
\EndProcedure
\end{algorithmic}
\end{algorithm}

\subsection{Active Defense}
We describe how forwarding accountability serves as a building block for
active DDoS defense. Transit ISPs can simply drop traffic from malicious
ASes, providing a primitive DDoS defense. However, accountable
proof of misbehavior can be combined seamlessly with more sophisticated
protection schemes.

Filtering defense proposals (e.g., StopIt~\cite{stopit}, AITF~\cite{aitf},
and Pushback~\cite{pushback}) demonstrate the effectiveness of a
distributed and cooperative approach to control certain traffic flows by
asking upstream routers to install filters. These approaches assume that
upstream routers are willing to install such filters. However, at the
inter-domain level this is a strong assumption.

ISPs are harsh competitors and are mutually distrusted entities. In
addition, ISPs earn revenue by forwarding traffic, regardless of the intent
of the traffic. Furthermore, filtering resources at forwarding devices are
limited and should be used cautiously. Hence, spending filtering resources
for targets outside the AS boundaries is an assumption that does not hold.
StopIt~\cite{stopit} recognizes this fact for inter-domain filtering
requests and leverages shared keys to authenticate such requests. However,
no filtering proposal obtains proof of misbehavior in order to
install such filters. Malicious ASes could try to exhaust filtering
resources of other ASes. 

\name allows an AS to provide misbehavior proof to
other ASes and convince them to install filters. Furthermore,
accountability can lead to novel contractual regimes and SLAs that formally
describe cooperative mechanisms to address the flooding attacks.

% Discussion
\label{sec:disc}
We discuss the deployment and operation of \name. 
The prominent advantage of \name is founded on the fact that
collateral damage can be leveraged to push ISPs to enforce higher
security standards, e.g., to deal with internal security threats such
as botnets or vulnerable components. Collateral damage mainly stems
from today's Internet architecture, and specifically from its lack of
accountability. In particular, in distributed attacks, the
misbehaving source end hosts cannot be identified.

\name identifies such malicious sources at the AS granularity with
the consequence that also innocent flows get classified as malicious.
Clearly, harming innocent flows is undesirable, but provable AS
misbehavior gives incentives for ISPs to take action against such
malicious traffic (e.g., deprioritize or drop it). This holds the
whole AS accountable for misbehavior and puts it under pressure to
deal with its security issues, rather than delegating flooding
protection to the victim. Hence, provable misbehavior turns
collateral damage on its head by using innocent flows as a way to
pressure ASes to deal with their security issues.

\subsection{Deployment Path} 

\name is deployable in the context of today's Internet as it does not
require architectural changes. More precisely, \name is
compatible with today's protocols and especially with IPv6
extension headers, which were designed for deploying 
novel protocols. The introduced overhead, although not
negligible, is within reach of today's processing and networking
capabilities. In addition, given that source and destination ASes set
up a sending policy, the destination can protest and prove
misbehavior even if only one transit AS supports \name. Thus, ASes
can deploy \name independently without global coordination.

On the downside, forwarding devices on the data path will need to
support additional processing mechanisms, which translate to
upgrades and costs. Furthermore, the considerable storage overhead
for destination ASes can further increase operational costs.
Finally, the requirement for a policy construction that defines the
characteristics of the transmission constitutes a deviation from today's
communication model.

\subsection{Operational Assumptions} In the high-level
overview of \name (\autoref{sec:overview}), we presented a
router-level communication model between the source and destination
AS in which we assumed that all traffic flows originated by the
source AS follow the same AS-level path towards the destination. We
relax this assumption of a line topology, as this model does not
reflect reality: each border router decides independently on the next
AS hop. Moreover, the interaction of inter-domain routing and
intra-domain traffic engineering (e.g., load balancing) leads to
different AS-level paths between the source and destination ASes. Therefore,
in \name, a communication channel is identified by the AS path and
not by the source-destination AS tuple. 

Furthermore, two ASes can peer at multiple Points of Presence. Consequently,
the source AS might have to coordinate the sending rates if there are
multiple peering points with the next AS. Readily available approaches
deal with such traffic engineering tasks: Segment Routing Centralized Egress
Peer Engineering developed by Cisco~\cite{segment_routing} and Intelligent Route Service 
Control Point solutions~\cite{irscp} are such examples.

Routing instability that forces source and destination to reestablish a
communication channel over a new path is not a notable concern. Studies show
that the majority of network routes are stable from tens of minutes to
days~\cite{route_stability, crossfire}. Despite ISPs' traffic engineering and
the existence of short-lived routes, long-lived routes are used 96\% of the
time~\cite{route_stability}.

Furthermore, today's border routers are not required to perform
cryptographic operations on data-plane traffic. However, the recent
advances in cryptographic engines, such as Intel AES-NI~\cite{aes_ni},
allow efficient cryptographic operations even for commodity machines,
as we have demonstrated in \autoref{subsec:eval}.

Moreover, schemes that increase the packet length (the border router
of the source AS adds the \name header) need to take into account 
correct MTU discovery. In case a large packet requires fragmentation,
the border router of the source AS can respond with an MTU size small enough,
so that the \name header can be added without concerns.

\subsection{Security Concerns} In this paper, we focused on the
security properties of the accountability framework and not on other
security aspects (such as source accountability or flooding attacks
on the channel setup). Source address spoofing is a well-known and
studied problem with best current security practices (BCP
38/84~\cite{rfc2827,rfc3704}) that should be followed by
administrators. 
Denial-of-Capability (DoC) attacks -- flooding the request channel of
capability defense systems -- have been demonstrated along with
proposals for defense~\cite{tva,portcullis}, which can be used as
protection from flooding the \name setup channel. We stress that  
our key ideas are compatible with other future Internet proposals
that address natively the aforementioned security concerns~\cite{scion_arxiv,scion}.

\section{Related Work}
\label{sec:related}
We describe some major accountability and DDoS defense
schemes; comprehensive surveys about DDoS defense can be
found in Zargar et al.~\cite{ddos_survey} and in Mirkovic et
al.~\cite{mirkovic_book}.

Accountability mechanisms are building blocks to hinder DDoS
attacks, rather than active defense mechanisms. For example,
\textbf{AIP}~\cite{aip} is a network architecture based on accountability,
with a two-level flat
addressing structure that allows for using self-certifying addresses (the
hash of the corresponding public keys). \textbf{IPA}~\cite{ipa} is a more
lightweight approach that binds an IP prefix to the public key of an
AS by leveraging the DNSSEC infrastructure. The secured bindings are 
piggybacked in BGP messages and get distributed in a protocol-compliant
and incrementally-deployed way. \textbf{Passport}~\cite{passport}
is a network-layer source authentication system that authenticates the
source of a packet to the granularity of the origin AS. Symmetric key
cryptography is used and packets are checked only at administrative
boundaries.  Using accountable source addresses as a building block,
additional defense schemes are proposed. For example, a shut-off protocol
is proposed~\cite{aip}, where a host can instruct the network interface of
an attacker to stop packet transmission. However, this pushes DDoS defense
to the hosts, assuming that all hosts recognize such a shut-off protocol.

Simon et al.\ propose AS-based accountability as a cost-effective DDoS defense~\cite{as_acc}.
Moreover, the authors propose an evil bit in the packet headers. The
proposal works for a group of participating ASes, assuming pairwise and transitive
trust between them. The evil bit is set whenever traffic enters from outside the island
of the participating ASes. However, the inferred threat model is weak, since a single compromised
AS inside the group of participating ASes limits the effectiveness of the proposal. In addition,
the system introduces considerable upgrades in terms of infrastructure and requires new Customer
Relationship Management (CRM) systems.

Other accountability schemes used for debugging and forensics
introduce prohibitive overhead for deployment in the data plane.
SNP~\cite{snp}, PeerReview~\cite{peerreview}, and \mbox{NetReview}~\cite{netreview}
keep detailed logs of exchanged messages and introduce substantial
overhead in terms of processing, storage, and bandwidth.

An alternative approach to identify the source of an attack is to identify the path(s)
traversed by malicious traffic. In \textbf{IP
traceback}~\cite{traceback}, downstream routers probabilistically
mark packets with partial path information. The victims combine the
partial path information in the packets to reconstruct the path(s) to
the source(s) of the attack. The proposal yields high computational
overhead for path reconstruction at the victims and a high false
positive rate even for small scale DDoS attacks~\cite{sec_traceback}.
In addition, IP traceback operates under a weak threat model, in which
downstream routers need to be trusted. Incremental proposals optimize
the computational overhead and operate under a stronger threat model
that includes malicious routers~\cite{sec_traceback}. \textbf{Hop-Count
Filtering}~\cite{hcf} is a host-based approach that discards spoofed
DDoS traffic. The main idea is that the only IP header information that
cannot be influenced by an attacker is the TTL field. Hence, spoofed IP
packets will most probably have inconsistent hop-count values with the
IP addresses being spoofed. \name borrows ideas from these schemes, as
the packets contain proofs of misbehavior if the source violates
the acknowledged traffic profile.
The destination then sends the proofs
\textit{back} to the corresponding ASes to prove the misbehavior.

There are two main approaches for \emph{active} defenses against DDoS
attacks: capabilities and filtering. Capability
proposals~\cite{ticketing,siff,tva,portcullis} let the
destination explicitly authorize traffic that it desires to receive.
Our approach is inspired by capability schemes --- not
for proving traffic legitimacy, but for collecting and providing proofs to each
transit AS on the path. The first challenge for a
victim is to distinguish between malicious and benign traffic
sources~\cite{ticketing}. Benign traffic sources get short-term
authorizations -- capabilities -- from the destinations and put them
into the packets, so that the legitimacy of traffic can be verified.
Capability proposals introduce considerable complexity and are
susceptible to DoC attacks~\cite{siff}. To address DoC, \textbf{TVA}~\cite{tva} tags each packet with
the identifier of the ingress point to an AS and fair-queues packets
at each router according to this identifier.  
\textbf{Portcullis}~\cite{portcullis} uses puzzles
(computational proofs of work) to provide fair sharing of the request
channel. \textbf{NetFence}~\cite{netfence} is a hybrid system and introduces a secure
congestion policy feedback combined with elements from capability-based
systems. Most capability proposals assume a mechanism that
distinguishes malicious from benign traffic and the effectiveness of
these proposals is, at most, as good as this assumed mechanism. 
In \name, we use a traffic profile that draws a clear line between
malicious and benign behavior, and use the proofs in the packets
to push the edge ASes to address their security problems.

The second class of active DDoS defense mechanisms, filtering
proposals, relies on stopping malicious flows in the network before
reaching the victim. \textbf{StopIt}~\cite{stopit} uses a
closed-control and open-service architecture to defend from attacks
that prevent filter installation. End hosts can send StopIt requests
only to their access routers and each AS has a StopIt server that
handles StopIt requests.  \textbf{AITF}~\cite{aitf} installs filters in
routers as close as possible to the attacking sources, rather than in
backbone routers.  \textbf{Pushback}~\cite{pushback} detects a
malicious traffic aggregate and controls it at a single router and in a
cooperative manner by asking upstream providers to stop the malicious
aggregate.  Such filtering schemes assume cooperation among ISPs and
that ISPs are willing to provide some of their filtering resources to
protect remote victims. However, this is an unrealistic assumption in
today's competitive Internet ecosystem and we consider the accountable
proof of misbehavior as a way to convince ISPs to install filters.
Alternatively, such proof can lead to new contracts among ISPs with
regard to security.

\section{Conclusion}
\label{sec:conc}
This paper has presented \name, an attempt to answer the question on
how to incentivize ISPs to adopt stricter security policies and
thereby to secure the insecure edge of the Internet where most of
today's security problems are rooted.

\name leverages forwarding accountability to prove to transit ISPs on
the path from the source to the destination that they have forwarded
(malicious) traffic. Using \name's accountable proof of misbehavior, we
have presented an application -- the suspicious bit -- that
incentivizes ISPs to mark traffic from their suspicious customers as
such and thereby inform other entities in the network.
\name\ comes with less than 2\% bandwidth overhead and without any
storage overhead for the transit ISPs.
Furthermore, \name is incrementally deployable in today's Internet,
and it gives incentives for early adoption.

We have implemented a \name software switch that processes packets at
the line rate of 120~Gbps, and forwards 140M minimum-sized packets per
second.

\newcommand{\BIBdecl}{\setlength{\itemsep}{0.8em}}
\bibliographystyle{myIEEEtran}
\bibliography{macros.long.bib,fullbib,rfc}

\end{document}